\documentstyle[twoside,fleqn,espcrc2,epsf]{article}

\def\spose#1{\hbox to 0pt{#1\hss}}
\def\ltapprox{\mathrel{\spose{\lower 3pt\hbox{$\mathchar"218$}}
 \raise 2.0pt\hbox{$\mathchar"13C$}}}
\def\gtapprox{\mathrel{\spose{\lower 3pt\hbox{$\mathchar"218$}}
 \raise 2.0pt\hbox{$\mathchar"13E$}}}
\def\inapprox{\mathrel{\spose{\lower 3pt\hbox{$\mathchar"218$}}
 \raise 2.0pt\hbox{$\mathchar"232$}}}

\epsfverbosetrue

\newcommand{\ewxy}[2]{\setlength{\epsfxsize}{#2}\epsfbox[10 30 640 590]{#1}}

\newcommand{\plus}{\makebox[15pt][c]{$+$}}
\newcommand{\minus}{\makebox[15pt][c]{$-$}}
\newcommand{\ds}{\displaystyle}
\newcommand{\be}{\begin{equation}}
\newcommand{\nn}{\nonumber}
\newcommand{\ee}{\end{equation}}
\newcommand{\bea}{\begin{eqnarray}}
\newcommand{\bat}{\begin{array}{cc}}
\newcommand{\ba}{\begin{array}}
\newcommand{\ea}{\end{array}}

\newcommand{\eea}{\end{eqnarray}}

\newcommand{\err}[2]{\raisebox{0.08em}{\scriptsize {$\;\begin{array}{@{}l@{}}
                          \plus\makebox[0.55em][r]{#1} \\[-0.12em] 
                          \minus\makebox[0.55em][r]{#2} 
                        \end{array}$}}}
\newcommand{\er}[2]{\raisebox{0.08em}{\scriptsize {$\;\begin{array}{@{}l@{}}
                          \plus\makebox[0.15em][r]{#1} \\[-0.12em] 
                          \minus\makebox[0.15em][r]{#2} 
                        \end{array}$}}}

\newcommand{\pslash}{\rlap{p}{\kern0.1em\hbox{/}}}
\newcommand{\vslash}{\rlap{v}{\kern0.07em\hbox{/}}}
\newcommand{\Dslash}{\rlap{D}{\kern0.1em\hbox{/}}}
\newcommand{\qslash}{\rlap{q}{\kern0.1em\hbox{/}}}
\newcommand{\ff}{Form Factors }

\def\Mb{M_{\Lambda_b}}
\def\Mc{M_{\Lambda_c}}

\def\Mb{M_{\Lambda_b}}
\def\Mc{M_{\Lambda_c}}

\title{A first study of the semi-leptonic decay of the $\Lambda_b$ baryon.
{\small \hfill SHEP prep. 97/17}}

\author{
Nicoletta Stella\address{
Physics Department, University of Southampton, SO17 1BJ
Southampton, UK} {\em for the UKQCD Collaboration}
}

\begin{document}
\begin{abstract}
We present the preliminary results of the first Lattice study of the baryonic
Isgur and Wise function obtained from the matrix element of the weak current between
$\Lambda$-baryon external states. Its dependence on the heavy and light quark
masses is studied. Some result on the semi-leptonic decay $\Lambda_b\to\Lambda_c 
+l\nu$ are given.
\end{abstract}

\maketitle

We present the results of the first non-perturbative
study of the semi-leptonic decay $\Lambda_b \to \Lambda_c +l\nu $,
carried out using Lattice QCD.
This study will provide an independent measurement of the
CKM matrix element $V_{cb}$, as experimental data become available.

We evaluate the hadronic matrix element of the weak current $J_{\mu}=\bar c
\gamma_{\mu}(1-\gamma_5)b$, computing the correlators

\vspace{-0.2cm}

\be
C(t_x)\!=\!\!\!\ds{\sum_{\vec{x}}} 
 e^{-i\vec{p}\cdot\vec{x}} 
 \langle {\cal O}^{Q}(x) 
 \bar{\cal O}^{Q}(0) \rangle 
\label{prima}
\ee

\vspace{-0.5cm}

\[
C_{\mu}(t_x,t_y)\!=\!\!\!\ds{\sum_{\vec{x},\vec{y}}}
 e^{-i(\vec{p}\cdot\vec{x}+\vec{q}\cdot\vec{y})} 
 \langle {{\cal O}}^{Q}(x) 
 (J_{\mu}(y))
 \bar{{\cal O}}^{{Q}}(0) \rangle
\]

\vspace{-0.2cm}
\noindent on 60 $24^3\times 48$ lattices 
at $\beta=6.2$, and using the $O(a)-$improved fermion action \cite{SW}.
In (\ref{prima}) $ {\cal O}^Q$ is the interpolating operator of the 
$\Lambda_Q$.
In this study we consider only correlators with initial and final momenta
either zero or 
$p_{\rm min}=2\pi/La$ and transitions with equal initial and final heavy 
quark masses $m_Q$.
The matrix element can be decomposed 
into six form factors (FF), which are invariant functions of $\omega=v\cdot v'$
\[
\langle \Lambda_Q(v')| J_{\mu} |
\Lambda_Q^{(r)}(v)\rangle\! =\! \bar{u}(v')
\Gamma_{i}(F^V_i-\gamma_5 F^A_i) u(v)\ .
\]
with $\Gamma_i=\gamma_{\mu},v_{\mu}$ and $v_{\mu}'$. 
In this basis, one can use the Heavy Quark Effective Theory (HQET) analysis
to relate the six FF to the Isgur-Wise (IW) function \cite{NEUBERT}, through the
correction coefficients $N_i^{(5)}$
\[
N_i(\omega)\hat\xi_{QQ'}(\omega)\! =\! F^V_i(\omega)\ \ \
N_i^5(\omega)\hat\xi_{QQ'}(\omega)\! =\! F^A_i(\omega). \nn
\]
$\hat\xi_{QQ'}(\omega)$ is explicitly flavour-dependent, but still
normalized to 1 and protected 
by Luke's theorem at $\omega=1$.

We will study the quantities
\be
\hat\xi_{QQ'}(\omega)=
\frac{F^A_1(\omega)}{F^A_1(1)}\frac{N_1^5(1)}{N^5_1(\omega)}=
\frac{\sum_i F^V_i(\omega)}{\sum_i F^V_i(1)}
\frac{\sum_i N_i(1)}{\sum_i N_i(\omega)}\ \
\label{equality}
\ee
which are independent of the current renormalization constants,
and exhibit a stable signal. The equality
(\ref{equality}) is valid up to $O(1/m_Q)$, and we neglect
higher order corrections.
\begin{table*}
\begin{center}
\begin{tabular}{|l|r|c|c|c|c|c|c|c|c|}
\hline
$\kappa_{l1}/\kappa_{l2}$& $J_{\mu}$&$\omega$&$\hat\xi_{QQ'}$&
$\omega$&$\hat\xi_{QQ'}$&$\omega$&$\hat\xi_{QQ'}$&$\omega$&$\hat\xi_{QQ'}$\\
\hline
$0.14144/$&A&$1.037$&$0.94\er{3}{3}$&$1.08$&$0.88\er{9}{9}$&$1.0$&$1.02\err{11}{11}$&$1.15$&$0.68\er{7}{6}$\\
$0.14144 $&V&$1.037$&$0.96\er{3}{3}$&$1.08$&$0.90\er{8}{8}$&$1.0$&$0.98\err{11}{12}$&$1.15$&$0.62\er{8}{7}$\\
\hline
$0.14144/$&A&$1.040$&$0.93\er{3}{4}$&$1.08$&$0.86\err{12}{11}$&$1.0$&$0.97\err{15}{16}$&$1.16$&$0.71\er{9}{9}$\\
$0.14226 $&V&$1.040$&$0.97\er{3}{4}$&$1.08$&$0.90\err{10}{10}$&$1.0$&$0.86\err{16}{17}$&$1.16$&$0.65\er{9}{8}$\\
\hline
$0.14226/$&A&$1.043$&$0.93\er{5}{6}$&$1.09$&$0.82\err{18}{15}$&$1.0$&$0.85\err{24}{25}$&$1.18$&$0.75\err{13}{12}$\\
$0.14226 $&V&$1.043$&$0.95\er{5}{5}$&$1.09$&$0.80\err{13}{14}$&$1.0$&$0.63\err{27}{27}$&$1.18$&$0.68\err{13}{11}$\\
\hline
Chiral/& A&$1.048$&$0.92\er{7}{7}$&$1.10$&$0.90\err{22}{20}$&$1.0$&$1.02\err{23}{24}$&$1.19$&$0.78\err{17}{17}$\\
Chiral&V&$1.048$&$0.96\er{8}{9}$&$1.10$&$0.95\err{19}{19}$&$1.0$&$0.79\err{26}{32}$&$1.19$&$0.74\err{16}{15}$\\
\hline
\end{tabular}
\caption{Estimates of \protect$\hat\xi_{QQ'}(\omega)\protect$ form both axial and
vector form factors, at $\kappa_Q=\kappa_Q'=0.129$ and at three values of the light
quark masses and at the chiral limit. Errors on $\omega$ are on the
digit beyond the last shown.\label{tabextra}}
\end{center}
\vspace{-0.3cm}
\end{table*}

\begin{figure}
\begin{picture}(80,45)
\put(0,-13){\ewxy{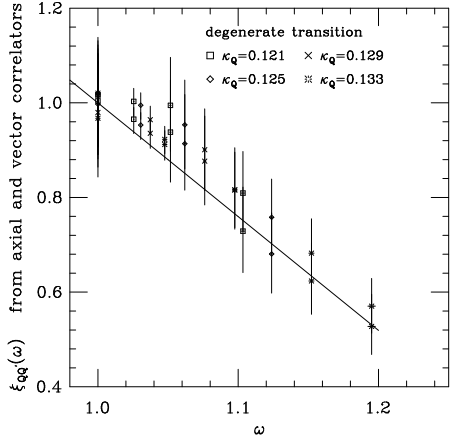}{80mm}}
\end{picture}
\caption{\protect$\hat\xi_{QQ'}(\omega)\protect$ from degenerate transitions, 
vector and axial current. Different graphical symbols denote different heavy
quarks.}
\label{figFG}
\vspace{-0.3cm}
\end{figure}

The explicit flavour-dependence of $\hat\xi_{QQ'}$ is
studied by linearizing 
it about $\omega=1$
\be
\hat\xi_{QQ'}(\omega)= 1 +\rho^2(1-\omega)
\label{fitfunc}
\ee
and measuring the slope $\rho^2$ as a function of $m_Q$.
By keeping the light quark masses fixed to $\kappa=0.14144$, around 
that of the strange quark, 
and varying $m_Q$  around the charm mass, we obtain
\bea
\rho^2= 2.4\er{4}{4}  &\ \ {\rm at }\  &\kappa_Q=\kappa_Q'=0.121 \nn \\
\rho^2= 2.4\er{4}{4}  &\ \ {\rm at }\ \ &\kappa_Q=\kappa_Q'=0.125 \nn \\
\rho^2= 2.4\er{3}{4}  &\ \ {\rm at }\ \ &\kappa_Q=\kappa_Q'=0.129 \nn \\
\rho^2= 2.4\er{3}{3}  &\ \ {\rm at }\ \ &\kappa_Q=\kappa_Q'=0.133 ,
\label{fourvalues}
\eea
suggesting that the flavour dependence of $\hat\xi_{QQ'}$ can
be neglected at our masses or above.
A global fit, shown in Figure~\ref{figFG}, to the four sets of 
determinations, yielding  $\rho^2=    2.4\er{4}{4}$,
is our best estimate.

According to HQET, the IW function
only depends on the quantum numbers of light quarks.
Previous studies on the lattice \cite{lpl} 
demonstrated that such a dependence is not negligible in the case of mesons,
where the ``brown muck''  contains only one light quark.
Thus we expect to measure an even stronger
dependence of the baryonic IW function.

This was studied fixing the heavy quark mass to
$\kappa_{Q}=\kappa_{Q'}=0.129$, i.e. very close to the charm quark,
and considering the three light hopping-parameter combinations
$(\kappa_{l1}=\kappa_{l2}=0.14144 )$,$(\kappa_{l1}=0.14226,
\kappa_{l2}=0.14144)$, and
$(\kappa_{l1}=\kappa_{l2}=0.14226 )$.
By fitting together the quantities (\ref{equality}) to
Eqn.~(\ref{fitfunc}), it was found:
\bea
\rho^2=2.4\er{4}{4} &\ {\rm at}\ &\kappa_{l1,2}=0.14144, 0.14144\nn\\
\rho^2=2.0\er{5}{5} &\ {\rm at}\  &\kappa_{l1,2}=0.14144, 0.14226\nn\\
\rho^2=1.7\er{6}{8} &\ {\rm at}\  &\kappa_{l1,2}=0.14226, 0.14226
\label{threelight}
\eea
The slope $\rho^2$ at the chiral limit, 
\be
\rho^2=1.2\err{8}{11},
\ee
is obtained by extrapolating the three estimates of 
both $\omega$ and $\hat\xi_{QQ'}(\omega)$ 
linearly  in the sum of the two light quark masses,
Our results are presented  in Table~\ref{tabextra}. 

Our estimate of the IW function can, in turn, be used to quantify the 
$1/m_Q$ corrections, affecting $F_1$ and $F_{2,3}$, in terms of
the baryonic binding energy $\bar\Lambda$, as detailed in \cite{noi2}.
There, it was found that $\bar\Lambda=0.37\pm 0.11$ GeV.
Its value is necessary to reconstruct the FF at the physical limit,
i.e. for the decay $\Lambda_b \to \Lambda_c + e \nu $.

With the result that $\hat\xi_{QQ'}(\omega)$ is flavour-independent
the FF depend on the quark masses only through the factors $N_1^{(5)}$.
Given our limited knowledge of the functional form of
$\hat\xi_{QQ'}(\omega)$, we
can only model the FF as linear functions of $\omega$
\be
F^{A,V}_i(\omega,m_b,m_c) = \eta_i^{A,V} -\tilde{\rho}^{A,V}_i(\omega-1) 
\ee
where the normalizations $\eta_i^{V,A}$ and the new slopes $\tilde\rho_i^{A,V}$
 are related to the 
coefficients $N_i^{(5)}(\omega,m_b,m_c)$ and to the slope of the IW function by
\[
\eta_i^{V,A}\!=\!N_i^{(5)}(1), \ \ \
\tilde{\rho}_i^{V,A}\!=\!\! \rho^2\eta_i^{V,A}-
\left.\ds{\frac{d N_i^{(5)}(\omega)}{d \omega}}\right|_{\omega=1}
\]
Our results for $\tilde\rho_i^{V,A}$ and $\eta_i^{V,A}$ are shown in Table~\ref{tab:newslope}.

\begin{table*}
\begin{center}
\begin{tabular}{|c|cccccc|}
\hline
-&$F^V_1$&$F^V_2$&$F^V_3$&$F^A_1$&$F^A_2$&$F^A_3$\\
\hline
$\tilde\rho$&$1.8\err{9}{15}$&$-0.4\er{2}{1}$&$-0.10\er{7}{4}$&
      $1.3\err{8}{12}$&$-0.4\er{3}{2}$&$0.16\err{6}{10}$\\
$\eta$&$1.28\er{6}{6}$&$-0.19\er{4}{4}$&$-0.06\er{2}{1}$&$0.99$&$-0.24\er{5}{4}$&$0.09\er{2}{2}$\\
\hline
\end{tabular}
\end{center}
\caption{\em Normalization and slope of the physical \ff, relevant for the $
\Lambda_b$ decay. \label{tab:newslope}}
\vspace{-0.3cm}
\end{table*}

The decay rates can be written in terms of the FF, in the velocity basis,
through the helicity amplitudes $H$ \cite{DESY}:
\bea
\frac{d\Gamma}{d\omega} = \frac{G_F^2}{(2\pi)^3} 2|V_{cb}|^2
2\frac{q^2 \Mc^2\sqrt{(\omega^2-1)}}{12 \Mb}\times\qquad\qquad\qquad &&
\hspace{-1.7cm} (9) \nn
\\
\left( 
|H_{1/2,1}|^2 + |H_{-1/2,-1}|^2 +
|H_{1/2,0}|^2 +  |H_{-1/2,0}|^2  \right) ,&&\nn
\eea
where the subscripts 
in the helicity amplitudes refer to the
polarization of the $W$ boson and of the daugther baryon $\Lambda_c$, respectively.
The upper integration limit on the decay rates extends to 
$\omega\simeq 1.43$, which is beyond 
the range of velocity transfer accessible to
us ($\omega\in[1.0,1.2]$). We thus define the 
partially-integrated decay rate, 
\be
\Gamma^{\rm part}(\omega_{\rm max}) =\int_1^{\omega_{\rm max}} d\omega \frac{d \Gamma}{d\omega}
\nn
\ee
as a function of the upper limit of integration.
In Table~\ref{tabrate}, we present our results for the quantities
\be
\frac{\Gamma^{\rm part}(\omega_{\rm max})}{|V_{cb}|^2 }[10^{13} s^-1]
\ee
for several values of $\omega_{\rm max}$. The masses are taken from the
experiments.

\begin{figure}
\begin{picture}(80,45)
\put(0,-13){\ewxy{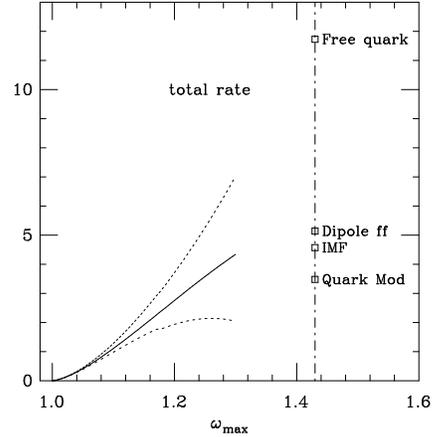}{80mm}}
\end{picture}
\caption{\em Total decay rate with error bands, for the process $\Lambda_b\to\Lambda_c + l\nu$, 
as a function of the limit of integration 
$\omega_{\rm max}$. A comparison with several model estimates is shown at the end-point.}
\label{fig:rate}
\vspace{-0.3cm}
\end{figure}

At present, a direct comparison of our results with experiments is not possible.
In fact, even if the semi-leptonic decay of $\Lambda_b$ was observed by various 
experiments \cite{LEPSEMIL}, a measurement of the decay rate is not yet available.
The problem of determining the rate of the $\Lambda_b$ semi-leptonic
decays has been addressed making use of different models (Infinite Momentum Frame,
Quark Model, Dipole form factors ) \cite{DESY},\cite{ivanov}. Their predictions, for the total rate, integrated up to the end-point,
are reported in Figure~\ref{fig:rate}, and compared with the function
$\Gamma^{\rm part}(\omega_{\rm max})$. To evaluate this function we
have assumed $|V_{cb}|=0.044$.

\begin{table}
\begin{tabular}{|l|cccc|}
\hline
$\omega_{\rm max}$&1.1& 1.15&1.20 &1.25  \\
$\Gamma_{\rm tot}^{\rm part}$&$ 0.57\er{9}{7}$&
$ 1.03\err{22}{20}$&$ 1.4\er{5}{4}$&$ 1.8\err{9}{7} $\\
\hline
\end{tabular}
\caption{\em Partial decay rates for the $\Lambda_b$. \label{tabrate}}
\vspace{-0.7cm}
\end{table}

Finally, we note that many other interesting quantities, 
such as asymmetry parameters
(see for example \cite{Nm2})
and the ratio of the longitudinal to transverse rates,
could be computed and confronted with the upcoming experiments. However,
all these quantities are non-trivial only at $O(1/m_Q^2)$ or
$O(\omega^2)$: in both cases beyond
the precision reached in the present study.


\def \ajp#1#2#3{Am. J. Phys. {\bf#1}, #2 (#3)}
\def \apny#1#2#3{Ann. Phys. (N.Y.) {\bf#1}, #2 (#3)}
\def \app#1#2#3{Acta Phys. Polonica {\bf#1}, #2 (#3)}
\def \arnps#1#2#3{Ann. Rev. Nucl. Part. Sci. {\bf#1}, #2 (#3)}
\def \cmts#1#2#3{Comments on Nucl. Part. Phys. {\bf#1}, #2 (#3)}
\def \cn{Collaboration}
\def \cp89{{\it CP Violation,} edited by C. Jarlskog (World Scientific,
Singapore, 1989)}
\def \efi{Enrico Fermi Institute Report No. EFI}
\def \f79{{\it Proceedings of the 1979 International Symposium on Lepton and
Photon Interactions at High Energies,} Fermilab, August 23-29, 1979, ed. by
T. B. W. Kirk and H. D. I. Abarbanel (Fermi National Accelerator Laboratory,
Batavia, IL, 1979}
\def \hb87{{\it Proceeding of the 1987 International Symposium on Lepton and
Photon Interactions at High Energies,} Hamburg, 1987, ed. by W. Bartel
and R. R\"uckl (Nucl. Phys. B, Proc. Suppl., vol. 3) (North-Holland,
Amsterdam, 1988)}
\def \ib{{\it ibid.}~}
\def \ibj#1#2#3{~{\bf#1}, #2 (#3)}
\def \ichep72{{\it Proceedings of the XVI International Conference on High
Energy Physics}, Chicago and Batavia, Illinois, Sept. 6 -- 13, 1972,
edited by J. D. Jackson, A. Roberts, and R. Donaldson (Fermilab, Batavia,
IL, 1972)}
\def \ijmpa#1#2#3{Int. J. Mod. Phys. A {\bf#1}, #2 (#3)}
\def \ite{{\it et al.}}
\def \jpb#1#2#3{J.~Phys.~B~{\bf#1}, #2 (#3)}
\def \lkl87{{\it Selected Topics in Electroweak Interactions} (Proceedings of
the Second Lake Louise Institute on New Frontiers in Particle Physics, 15 --
21 February, 1987), edited by J. M. Cameron \ite~(World Scientific, Singapore,
1987)}
\def \ky85{{\it Proceedings of the International Symposium on Lepton and
Photon Interactions at High Energy,} Kyoto, Aug.~19-24, 1985, edited by M.
Konuma and K. Takahashi (Kyoto Univ., Kyoto, 1985)}
\def \mpla#1#2#3{Mod. Phys. Lett. A {\bf#1}, #2 (#3)}
\def \nc#1#2#3{Nuovo Cim. {\bf#1}, #2 (#3)}
\def \np#1#2#3{Nucl. Phys. {\bf#1}, #2 (#3)}
\def \PDG{Particle Data Group, L. Montanet \ite, \prd{50}{1174}{1994}}
\def \pisma#1#2#3#4{Pis'ma Zh. Eksp. Teor. Fiz. {\bf#1}, #2 (#3) [JETP Lett.
{\bf#1}, #4 (#3)]}
\def \pl#1#2#3{Phys. Lett. {\bf#1}, #2 (#3)}
\def \pla#1#2#3{Phys. Lett. A {\bf#1}, #2 (#3)}
\def \plb#1#2#3{Phys. Lett. B {\bf#1}, #2 (#3)}
\def \pr#1#2#3{Phys. Rev. {\bf#1}, #2 (#3)}
\def \prc#1#2#3{Phys. Rev. C {\bf#1}, #2 (#3)}
\def \prd#1#2#3{Phys. Rev. D {\bf#1}, #2 (#3)}
\def \prl#1#2#3{Phys. Rev. Lett. {\bf#1}, #2 (#3)}
\def \prp#1#2#3{Phys. Rep. {\bf#1}, #2 (#3)}
\def \ptp#1#2#3{Prog. Theor. Phys. {\bf#1}, #2 (#3)}
\def \rmp#1#2#3{Rev. Mod. Phys. {\bf#1}, #2 (#3)}
\def \rp#1{~~~~~\ldots\ldots{\rm rp~}{#1}~~~~~}
\def \si90{25th International Conference on High Energy Physics, Singapore,
Aug. 2-8, 1990}
\def \slc87{{\it Proceedings of the Salt Lake City Meeting} (Division of
Particles and Fields, American Physical Society, Salt Lake City, Utah, 1987),
ed. by C. DeTar and J. S. Ball (World Scientific, Singapore, 1987)}
\def \slac89{{\it Proceedings of the XIVth International Symposium on
Lepton and Photon Interactions,} Stanford, California, 1989, edited by M.
Riordan (World Scientific, Singapore, 1990)}
\def \smass82{{\it Proceedings of the 1982 DPF Summer Study on Elementary
Particle Physics and Future Facilities}, Snowmass, Colorado, edited by R.
Donaldson, R. Gustafson, and F. Paige (World Scientific, Singapore, 1982)}
\def \smass90{{\it Research Directions for the Decade} (Proceedings of the
1990 Summer Study on High Energy Physics, June 25--July 13, Snowmass,
Colorado),
edited by E. L. Berger (World Scientific, Singapore, 1992)}
\def \tasi90{{\it Testing the Standard Model} (Proceedings of the 1990
Theoretical Advanced Study Institute in Elementary Particle Physics, Boulder,
Colorado, 3--27 June, 1990), edited by M. Cveti\v{c} and P. Langacker
(World Scientific, Singapore, 1991)}
\def \yaf#1#2#3#4{Yad. Fiz. {\bf#1}, #2 (#3) [Sov. J. Nucl. Phys. {\bf #1},
#4 (#3)]}
\def \zhetf#1#2#3#4#5#6{Zh. Eksp. Teor. Fiz. {\bf #1}, #2 (#3) [Sov. Phys. -
JETP {\bf #4}, #5 (#6)]}
\def \zpc#1#2#3{Zeit. Phys. C {\bf#1}, #2 (#3)}
\def \zpd#1#2#3{Zeit. Phys. D {\bf#1}, #2 (#3)}

\end{document}